\documentclass[11pt]{article}
\usepackage[utf8]{inputenc}

\usepackage{harrydoc}
\usepackage{harrymacros}

\DeclareMathOperator{\Inf}{Inf}
\DeclareMathOperator{\MaxInf}{MaxInf}
\DeclareMathOperator{\argmax}{argmax}

\usepackage[
    backend=biber,
    style=alphabetic,
    sorting=ynt,
    backref=true,
    maxbibnames=9,
]{biblatex}

\title{Small Shadow Partitions}

\newtoggle{blind}

\togglefalse{blind}

\toggletrue{showtodos}

\author{
    \iftoggle{blind}{}{
        Swastik Kopparty\thanks{Department of Mathematics and Department of Computer Science, University of Toronto.
        Research supported by an NSERC Discovery Grant.
        Email: swastik.kopparty@utoronto.ca} \and
        Harry Sha\thanks{Department of Computer Science, University of Toronto. Email: shaharry@cs.toronto.edu}
    }
}
\date{\today}

\newcommand{\mpv}[2]{\langle #1 \rangle_{#2}}

\begin{document}

\maketitle

\begin{abstract}
    We study the problem of partitioning the unit cube $[0,1]^n$ into $c$ parts so that each $d$-dimensional axis-parallel projection has small volume.

    This natural combinatorial/geometric question was first studied by Kopparty and Nagargoje \cite{KN} as a reformulation of the problem of determining the achievable parameters for seedless multimergers -- which extract randomness from ``$d$-where''' random sources (generalizing somewhere random sources). 
    This question is closely related to influences of variables and is about a partition analogue of Shearer's lemma.

    Our main result answers a question of \cite{KN}: for $d = n-1$, we show that for $c$ even as large as $2^{o(n)}$, it is possible to partition $[0,1]^n$ into $c$ parts so that every $n-1$-dimensional axis-parallel projection has volume at most $(1/c) ( 1 + o(1) )$. 
    Previously, this was shown by \cite{KN} for $c$ up to $O(\sqrt{n})$. 
    The construction of our partition is related to influences of functions, and we present a clean geometric/combinatorial conjecture about this partitioning problem that would imply the KKL theorem on influences of Boolean functions.
\end{abstract}

\tableofcontents

\section{Introduction}

In this paper, we study a basic combinatorial/geometric problem about partitioning product sets to minimize projections. 
This problem arose naturally in a work of Kopparty and Nagargoje~\cite{KN} on randomness extractors, and turns out to be closely related to the theory of influences of variables. 
Consider the solid $n$-dimensional unit cube $[0,1]^n$.
We want to partition this cube into $c$ parts\footnote{We assume the parts are ``nice'' so that there are no measure theoretic difficulties. 
Ultimately, all our results will based on a discrete analogue -- partitioning $[N]^n$ -- where such issues do not arise.} so that 
each part has all its $n-1$-dimensional axis-parallel projections\footnote{Note that there are $n$ different $n-1$-dimensional axis-parallel projections.} having volume at most $\beta$. 
The problem is to determine how $c$ and $\beta$ must be related for such a partition to exist.

Let $A_1, \ldots, A_c \subseteq [0,1]^n$ be a partition of $[0,1]^n$. 
For $i \in [n]$, let $\pi_{-i} : [0,1]^{n} \to [0,1]^{n-1}$ be the projection map onto all coordinates except the $i$'th.
For any fixed $i$, we clearly have that $\pi_{-i}(A_1), \ldots, \pi_{-i}(A_c)$ covers $[0,1]^{n-1}$; thus at least one of the $\pi_{-i}(A_j)$ has volume at least $\frac{1}{c}$. 
So we must have $\beta \geq \frac{1}{c}$.

We now highlight three important observations from~\cite{KN} that set the context for our main result.
\begin{enumerate}
 \item {\bf Near-perfect partitions:} When $c = 2$, the halfspace partition 
 $A_1 = \{ x \in [0,1]^n \mid \sum x_i \leq n/2 \}$, $A_2 = \{x \in [0,1]^n \mid \sum x_i > n/2 \}$  achieves $\beta = (1+ o(1))\frac{1}{2}$. Thus for each $i \in \{1, 2\}$, the sets
 $\pi_{-i}(A_j) $ form a near-disjoint cover of $[0,1]^{n-1}$.
 
 More generally, for $c = O(\sqrt{n})$, a similar construction
 with $A_i$ of the form $\{ x \in [0,1]^n \mid \sum x_i \in [t_i, t_{i+1}) \}$ achieves $\beta = (1 + o(1) ) \frac{1}{c}$.  
 \item {\bf Classical inequalities: } The classical inequalities of Shearer / Loomis-Whitney / Bollob\'as-Thomason give interesting information about individual parts of such a partition. Concretely, they say that any set with all $n-1$-dimensional axis-parallel projections have volume at most $\beta$, must have volume at most $\beta \cdot \beta^{1/(n-1)}$. 
 Since some part $A_i$ of the partition must have volume at least $\frac{1}{c}$, this implies a lower bound on $\beta$ of $c^{1/n} \cdot \frac{1}{c}$, which for $c = 2^{o(n)}$ is only $(1 + o(1)) \cdot \frac{1}{c}$, not much better than the trivial $\frac{1}{c}$ lower bound.
 
 \item {\bf Cube partitions:} Once $c$ gets exponentially large, the previous lower bound becomes a constant factor larger than $\frac{1}{c}$. 
 This turns out to be tight. Indeed, when $c = 2^n$, the trivial
 equipartition of $[0,1]^n$ into $2^n$ cubes of side length $1/2$ 
 is easily seen to have $\beta = \frac{2}{c}$.
\end{enumerate}

This raises the natural question about what happens when $c$ is in the intermediate range between $\sqrt{n}$ and $2^n$. Our main result is an answer to this question.

{\bf Main Theorem (Informal)}\ \ {\em For $c = 2^{o(n)}$, there exists a partition of $[0,1]^n$ into $c$ parts $A_1, \ldots, A_c$, such that for all parts $A_j$ and all $i \in [n]$, the $n-1$-dimensional axis-parallel projection $\pi_{-i}(A_j)$ has volume at most $(1 + o(1)) \frac{1}{c}$.}

Our construction of this partition is based on a simple connection between projections and influences of variables. 
Using this connection, we show how to use low-influence Boolean functions (such as Tribes) with a product construction to produce partitions whose parts have all their projections abnormally small.

For finer scale behavior, we propose the following conjecture.

{\bf Conjecture J (Informal)}\ \ {\em Let $b$ be a constant in $(0,1)$ and let $c = 2^{b n}$.
For any partition of $[0,1]^n$ into $c$ parts, some part has an $(n-1)$-dimensional axis-parallel projection of volume at least $(1 + \frac{1}{100} b \log \frac{1}{b}) \cdot \frac{1}{c}$.
}

If true, Conjecture $J$ would be tight upto the constant $\frac{1}{100}$; this follows from the construction that proves our Main Theorem above.

If we replace $b \log \frac{1}{b}$ in Conjecture J with $\Omega(b)$,
then this statement is true; it is directly implied by the Shearer/Loomis-Whitney/Bollob\'as-Thomason inequalities applied to the largest part.

Thus a proof of Conjecture J will have to truly use the fact that we have a partition of $[0,1]^n$, and not merely the fact that some part of the partition has volume at least $1/c$. 
Philosophically, this is the difference between Ramsey's theorem and Turan's theorem for the existence of cliques in graphs or between van der Waerden’s theorem and Szemeredi’s theorem for the existence of APs in sets of integers. Note that in~\cite{KN} this phenomenon was witnessed in
in the case $n=3$ and $c = 2$ and $3$.

We think it would be very interesting if Conjecture J is true.
Utilizing the same connection with influences that led to our Main Theorem, we observe that Conjecture J implies the KKL theorem from the theory of Boolean functions. 
Today the only known proof of the KKL theorem goes through the Fourier analysis of Boolean functions and hypercontractivity. 
We think it would be valuable to get a more geometric or combinatorial proof of the KKL theorem by perhaps directly proving the Conjecture J.

While the above discussion concerns $n-1$-dimensional projections, similar questions can be considered for $d$-dimensional projections for general $d$. We also make several observations about the more general setup.

The most interesting of these observations is what we call the Partition Sauer-Shelah lemma. 
The Sauer-Shelah lemma says that a large enough subset of $\{0,1\}^n$ will have a full projection $\{0,1\}^d$ when projected down to some $d$ coordinates. 
We consider the problem of determining for which $c$ and $d$ we have that every partition of $[0,1]^n$ into $c$ parts will have some part having some $d$-dimensional axis-parallel projection equalling $[0,1]^d$. 

{\bf Partition Sauer-Shelah:}\ \ {\em
Suppose $d \leq \left\lceil \frac{n}{c}\right\rceil$. 
Then for any partition $A_1, \ldots, A_c$ of $[0,1]^n$, there exists some part $A_j$ and some $d$-dimensional axis parallel projection $\pi_S(A_j)$ (where $|S| = d$) which equals $[0,1]^d$.

Conversely,
for $d > \left\lceil \frac{n}{c}\right\rceil$, then
there exists a partition
$A_1, \ldots, A_c$ of $[0,1]^n$, such that for all parts $A_j$, all $d$-dimensional axis parallel projection $\pi_S(A_j)$ (where $|S| = d$) have volume 
at most $1 - \epsilon_{c,d}$.
}

Understanding the quantitative behavior of $\epsilon_{c,d}$ and other related quantitative questions seems very interesting for further research, with possible connections to pseudorandomness and Boolean functions.

Finally we raise a closely related question that our method was not able to answer -- it is about a partition version of the Kruskal-Katona theorem. Let $N$ be huge as a function of $n$. Suppose we partition ${ [N] \choose n}$ into $c = n^{0.51}$ parts $\mathcal A_1, \ldots, \mathcal A_c$. Is it true that some part $\mathcal A_i$ must have shadow $\partial \mathcal A_i \subseteq { [N] \choose n-1}$ satisfying:
$$ | \partial \mathcal A_i| \geq (1 + \Omega(1) ) \cdot \frac{1}{c} \cdot { N \choose n-1 }?$$

\subsection{Other related work}

As mentioned before, these kinds of questions were first studied by
Kopparty and Nagargoje~\cite{KN} in the context of randomness extraction from ``$d$-where'' random sources. 
It was originally studied by~\cite{KN} in the context of partitions of $[N]^n$ for some integer $N$ much larger than $n$, but as noted there, it is equivalent to the analogous question for nice (without any measure-theoretic nastiness) partitions of $[0,1]^n$. Apart from the observations about large $n$ mentioned earlier,~\cite{KN} also found the optimal partition of $[0,1]^3$ into $2$ parts, showing that one of the two parts
must have a $2$-dimensional axis-parallel projection with area at least $3/4$.

In~\cite{CGR}, Chattopadhyay, Gurumukhani, and Ringach studied the problem of seedless condensing from a variety of weak random sources, generalizing $d$-where random sources. 
Their results gave a nearly complete picture on the kinds of NonOblivious Symbol Fixing (NOSF) sources (and variants) from which seedless condensing and extraction is possible.

The problem of minimizing projections also has connections to the BKKKL~\cite{bourgainInfluenceVariablesProduct1992,Friedgut} conjecture on Boolean functions on the solid cube $[0,1]^n$. 
This conjecture is still wide open, although there has been some interesting recent progress by Filmus, Hambardzumyan, Hatami, Hatami and Zuckerman~\cite{filmus2019biasing}.

Additionally, the setting of $d < n-s$ for $s > 1$, and $c=2$ is related to constructions of resilient functions - where coalitions of size $s$ have small influence. 
Ajtai and Linial show the existence of Boolean functions resilient to coalitions of size $n/\log^2(n)$ \cite{AL}, and Ivanov, Meka, and Viola find explicit constructions that nearly match \cite{IMV}. 
Also relevant is the randomized construction of Bourgain, Kahn, and Kalai \cite{BKK}. 
They show the existence of balanced Boolean functions $f$, for which every coalition of size at most $(1/2 - \delta)n$ can make the function equal to $1$ with probability at most $1 - n^{-C}$. 
In other words, the $(1/2 + \delta)n$-dimensional projections of $f^{-1}(1)$ have size at most $1 - n^{-C}$. 
The projection sizes of $f^{-1}(0)$, however, are not guaranteed to be small.

\subsection*{Organization of this paper}
In the next section, we set up some notation and define several key notions.
Our main result is proved in Section~\ref{sec:inf-upperbounds}, and can be read directly after some preliminaries in Section 2. 
Alternately, the reader is encouraged to read the few sections preceding it, which discuss the landscape of known results and some general observations about this problem at a leisurely pace.

\section{Preliminaries: partitions, projections, influences}

We will typically use $n$ to refer to the dimension of the hypercube to cover and $d$ to the dimension of the projections we are interested in (note $d \leq n$).  
A collection $\mathcal{C}$ of subsets of $\OIS^n$ is called a \emph{cover} of $\OIS^n$ if $\bigcup_{K \in \mathcal{C}} K = \OIS^n$. 
A cover of size $c$ is called a $c$-cover.
Elements $K \in \mathcal{C}$ are called \emph{parts}. 
If $\mathcal{C}$ is a cover of $\OIS^n$ and elements of $\mathcal{C}$ are disjoint, then call $\mathcal{C}$ a \emph{partition} of $\OIS^n$.
Another way to view $c$-partitions of $\OIS^n$ is as a function $f: \OIS^n \to [c]$, where the $i$th part of the partition is simply the preimage of $i$, $f^{-1}(i)$. 
We sometimes refer to elements of $[c]$ as colors and think of $f$ as assigning colors to elements of $\OIS^n$.

For any $A \subset [n]$, let $K_A$ be the projection onto the coordinates in $A$. I.e., the orthogonal projection onto the subspace spanned by $\{e_i: i \in A\}$.
Use $|\cdot|$ to denote the volume of $\cdot$ in the appropriate dimensions.
Define $$\mpv{\mathcal{C}}{n,d} = \max\left\{|K_A| : K \in \mathcal{C}, A \subset [n], |A| = d \right\}.$$
Refer to $\mpv{\mathcal{C}}{n, d}$ as the \emph{maximum $d$-dimensional projection volume of any part} of $\mathcal{C}$.
When a partition is specified by a function, $f$, let $\mpv{f}{n, d} = \mpv{\{f^{-1}(1),f^{-1}(2),...,f^{-1}(c)\}}{n,d}$
Define
$$
\rho_{n, d, c} = \min\left\{ \mpv{\mathcal{C}}{n, d}: \mathcal{C} \text{ is a } c \text{-cover of }\OIS^n \right\},
$$
and
$$
\kappa_{n, d, \epsilon} = \min\left\{ c \in \N : \exists \mathcal{C}. \mathcal{C} \text{ is a } c \text{-cover of }\OIS^n \text{ and } \mpv{\mathcal{C}}{n,d} \leq \epsilon \right\}.
$$

The two quantities are closely related, and proving bounds on one implies bounds on the other. 
Finding bounds on the $\rho_{n, d, c}$ amounts to answering the question of `what is the smallest possible $\mpv{\mathcal{C}}{n,d}$ of any $c$-cover $\mathcal{C}$?' On the other hand, finding bounds on $\kappa_{n, d, \epsilon}$ amounts to answering the question  `what is the smallest $c$ such that there is a $c$-cover $\mathcal{C}$ with $\mpv{\mathcal{C}}{n, d} \leq \epsilon$.
Depending on the setting, we may find it more convenient to work with $\kappa$ or $\rho$.

Let $f:\OIS^n \to [c]$, and $S \subset [c]$. 
Say $f$ is $\epsilon$-\emph{balanced} if $\Pr[f_1 = \alpha] \in [1/c - \epsilon, 1/c + \epsilon]$ for each $\alpha \in [c]$. Say $f$ is \emph{balanced} if $f$ is $0$-balanced.

Many of our partitions come from functions $f$ where the domain is the discrete cube $\OI^n$. 
Any function $f: \OI^n \to [c]$ can be extended to the domain $\OIS^n$ by rounding each coordinate.
In this view, each binary string $x \in \OI^n$ can be viewed as a hypercube of side length $1/2$.
For subsets of $K \subset \OI^n$, we sometimes use $|K|$ to denote the fractional size of $K$. 
This corresponds to the volume of $K$ when viewed as a subset of $\OIS^n$.

For any $\alpha \in [c]$, and $x \in \OI^S$, say $\alpha$ is \emph{above} $x$ if any of the following equivalent conditions are true.
\begin{enumerate}[1.)]
    \item There exists $y \in f^{-1}(\alpha)$ such that $y_S = x$.
    \item $x$ can be extended to some preimage of $\alpha$.
    \item $x \in f^{-1}(\alpha)|_S$.
\end{enumerate}
Let $A(x, S) = \{\alpha  \in [c]: \alpha \text{ is above x}\}$ be the set of colors above $x$.

The following is a useful characterization of the projection volumes from partitions $f: \OIS^n \to [c]$.
\begin{fact}
Let $f: \OIS^n \to [c]$, $S\subset [n]$, and $\alpha \in [c]$ then 
$$|f^{-1}(\alpha)_S| = \Pr_{x \in \OIS^{S}}[\alpha \text{ is above } x]$$
\end{fact}

\textbf{Influence.} For any $S \subset [n]$. The \emph{influence} of $S$, denoted $\Inf_f(S)$ is 
$$\Inf_f(S) = \Pr_{x \in \OIS^{\overline{S}}}[|A(x, \overline{S})| > 1].$$
Intuitively, it is the probability that one can change the output on a random input by changing bits restricted to $S$.
The \emph{maximum influence} of any subset of size $k$ is denoted $\MaxInf_k(f)$.
Abbreviate $\MaxInf_1$ as $\MaxInf$.

Next, we discuss two boolean-valued functions with low influence.

\textbf{Majority.} For any $n \in \N$, let $\Maj_{n} : \OI^n \to \OI$ be the majority function on $n$ bits. I.e. 

\[
    \Maj_{n}(x) = \begin{cases}
        1, &\text{ if }\wt(x) > n/2\\
        0, &\text{ else }
    \end{cases}
\]

It is well known that $\Maj_n$ is $O(1/\sqrt{n})$-balanced and has influence $O(1/\sqrt{n})$.

\textbf{Tribes.} The Tribes function was first presented in \cite{ben-orCollectiveCoinFlipping1985} and is defined as follows.
Let $n, w, s$ be integers such that $n = ws$. 
Break up $n$ input bits into $s$ `tribes' of size $w$ each.
$\Tribes_{w, s}: \OI^n \to \OI$ is the function which takes value 1 if and only some tribe is unanimously 1.
More formally, for any input $x \in \OI^n$, let $x^{(1)}, x^{(2)},...,x^{(s)}$ be sections of $x$ corresponding to each of the $s$ tribes (each $x^{(i)} \in \OI^w$), then
$$\Tribes_{w, s}(x) = \OR\left(\AND(x^{(1)}), \AND(x^{(2)}),...,\AND(x^{(s)})\right).$$

Depending on the relative sizes of $s$ and $w$, $\Tribes_{s, w}$ may be quite unbalanced.
\cite{ben-orCollectiveCoinFlipping1985} show how to choose the parameters such that $s, w$ is roughly balanced.
In particular, they show that there are infinitely many $n, s, w$ for which $\Tribes_{s,w}$ is $O(\log(n) / n)$-balanced, and the maximum influence is $\frac{\ln(n)}{n}(1 + o(1))$. Use $\Tribes_n$ to denote this particular setting of parameters.

Later, it will be important that the functions be completely balanced, so we adjust them slightly using the following fact.

\begin{fact}[Adjust Expectation.]
    Suppose $f: \OI^n \to \OI$ has maximum influence $\gamma$, and expectation $\mu$. Let $S \subset \OI^n$ with density $\epsilon$ such that $S \cap f^{-1}(1) = \emptyset$. Then, the function $f' = f + \1_S$ has expectation $\mu + \epsilon$, and maximum influence at most $\gamma + 2\epsilon$.
\end{fact}


Since for $\Maj_n$, and $\Tribes_n$, the balance and maximum influence are of the same order ($O(1/\sqrt{n})$ and $O(\log(n)/n)$, respectively) they can be modified to be fully balanced and have influence at most $O(1/\sqrt{n})$, and $O(\log(n)/n)$, respectively. For simplicity, in the rest of this paper, we will use $\Maj_n$ and $\Tribes_n$ to refer to the modified versions of the majority and tribes functions that are fully balanced.

Finally, we state the celebrated BKKKL theorem, which proves a lower bound on the influence of a functions from $f : \OIS^n \to \OI$.

\begin{theorem}[BKKKL \cite{bourgainInfluenceVariablesProduct1992}]\label{thm:bkkkl}
    Let $f: \OIS^n \to \OI$, such that $\Pr_{x \sim \OIS^n}[f(x) = 1] = p$, then 
    $\MaxInf(f) \geq \Omega(p(1-p)\log(n) / n).$
\end{theorem}

\section{Prior work}


\subsection{A general lower bound from classical projection inequalities}

We now give the general lower bound following the classical inequalities of Shearer/Loomis-Whitney/Bollob\'as-Thomason. 
The explicitly stated inequality that is most convenient for our presentation is the Uniform Cover Inequality of Bollob\'as and Thomason \cite{bollobasProjectionsBodiesHereditary1995}.

\begin{theorem}[Uniform Cover Inequality (\cite{bollobasProjectionsBodiesHereditary1995}, Theorem 2)]\label{thm:uniform-cover}
    Let $K \subset \OIS^n$, and $\mathcal{D}$ be a multiset of subsets of $[n]$ in which each $i \in [n]$ appears exactly $k$ times. Then,
    $$
    |K|^k \leq \prod_{A \in \mathcal{D}}|K_A|.
    $$
\end{theorem}

\begin{lemma}\label{lem:volume-lower-bound}
    Let $n, d$ be any positive integers such that $d \leq n$, and suppose $K \subset \OIS^n$.
    Then, there is some $A \subset [n]$ of size $d$ such that 
    $$|K_A| \geq |K|^{d/n}$$
\end{lemma}

\begin{proof}
Applying the Uniform Cover Inequality to $\mathcal{D} = \binom{[n]}{d}$ (and $k = \binom{n-1}{d-1}$), we get that for any $K \subset \OIS^n$, 
\begin{align*}
    |K|^{\binom{n-1}{d-1}} \leq \prod_{A \in \binom{[n]}{d}}|K_A|
\end{align*}
Let $A' = \argmax_{A \in \binom{[n]}{d}}(|K_A|)$, 
Then, we have 
$$
|K|^{\binom{n-1}{d-1}} \leq |K_{A'}|^{\binom{n}{d}}
$$
Hence, $|K_{A'}| \geq |K|^{d/n}$
\end{proof}

\begin{lemma}[General Lower Bound]\label{lem:general-lower-bound}
    Let $n, c, d$ be any positive integers such that $d \leq n$, and suppose $\mathcal{C}$ is a $c$-cover of $\OIS^n$.
    Then,
    $$\mpv{\mathcal{C}}{n,d} \geq (1 / c)^{d/n} \geq 1 - \frac{d \ln(c)}{n}$$
\end{lemma}

\begin{proof}
    Let $C$ be any cover of $\OIS^n$ of size $c$. 
    Then, at least one part $K \in \mathcal{C}$ has $|K| \geq 1 / c$.
    To get the first inequality, apply \Cref{lem:volume-lower-bound} to $K$.
    The second inequality follows from the following sequence of inequalities
    $$
    (1/c)^{d/n} = e^{-\frac{d\ln(c)}{n}} \geq 1 - \frac{d\ln(c)}{n}.
    $$
\end{proof}
As an immediate result, 
\begin{corollary}
    For any positive integers $n, d, c,$ with $d \leq n$, we have 
    $\rho_{n,d,c} \geq (1 / c)^{d/n}, $ and $ \kappa_{n, d, \epsilon} \geq (1/\epsilon)^{n/d}$.
\end{corollary}

Note that for $c = r^n$ for some $r \in \N$, The partition of $\OIS^n$ into $c$ hypercubes of side length $1/r$ is tight for this lower bound.



\subsection{Covering with solid hypercubes}
There is a line of work considering the covering of the $n$-dimensional torus, $[\R/\Z]^n$ using (solid) hypercubes of side length $\epsilon$.
Note that such a cover is also a covering of $\OIS^n$, where every part is a hypercube of side length $\epsilon$ that can potentially `wrap around at 1'.
Let $\mu_{n, \epsilon}$ be the minimum number of hypercubes of side length $\epsilon$ required to cover $[\R/\Z]^n$.
Since $d$-dimensional projections of a hypercube of side length $\epsilon$ is $\epsilon^d$, we have $\kappa_{n, d, \epsilon^d} \leq \mu_{n, \epsilon}$.

McEliece and Taylor \cite{mcelieceCoveringToriSquares1973} solve this problem for $n = 2$.
Let $\ceil{x}^{(i)} = \ceil{x \ceil{x}^{(i-1)}}$, and $\ceil{x}^{(0)} = 1$.

\begin{theorem}[\cite{mcelieceCoveringToriSquares1973}]\label{thm:d1-solution}
    For any $\epsilon \in (0,1)$, $\mu_{2, \epsilon} = \ceil{\epsilon^{-1}}^{(2)}$.
\end{theorem}

That is, they construct a covering of $[\R/\Z]^2$ with $\epsilon$-sided hypercubes of size $\ceil{\epsilon^{-1}\ceil{\epsilon^{-1}}}$, and prove a matching lower bound.
As it turns out, their lower bound technique can be adapted to more general coverings of $\OIS^n$ (where parts can be general shapes instead of only smaller hypercubes that can wrap around) and $d = 1$.

\begin{theorem}\label{lem:d1-lower-bound}
    Let $n, c$ be positive integers with $n\geq 1$, and let $\epsilon > 0$.
    If $\mathcal{C}$ is a $c$-cover of $\OIS^n$ with $\mpv{\mathcal{C}}{n,1} \leq \epsilon$, then $c \geq \ceil{\epsilon^{-1}}^{(n)}$.
    In other words, 
    $\kappa_{n, 1, \epsilon} \geq \ceil{\epsilon^{-1}}^{(n)}$.
\end{theorem}


We provide proof in \Cref{appendix:d1-lower} for completeness.
As a result, $\kappa_{2, 1, \epsilon} = \ceil{\epsilon^{-1}}^{(2)}$.
We translate this into $\rho_{2,1,c}$ and record the first few values below.
\begin{table}[h]
    \resizebox{\textwidth}{!}{
        \begin{tabular}{|l|l|l|l|l|l|l|l|l|l|l|l|l|l|l|l|}
        \hline
        $c$        & 1   & 2   & 3   & 4   & 5   & 6   & 7   & 8   & 9   & 10  & 11  & 12  & 13   & 14   & 15   \\ \hline
        $\rho_{2, 1, c}$ & 1/1 & 1/1 & 2/3 & 1/2 & 1/2 & 1/2 & 3/7 & 3/8 & 1/3 & 1/3 & 1/3 & 1/3 & 4/13 & 4/14 & 4/15 \\ \hline
        \end{tabular}
    }
\end{table}

Bogdanov, Grigoryan, and Zhukovskii \cite{bogdanovCoveringThreeToriCubes2021} extend this work to $n=3$, and find the optimal coverings of the three torus $[\R/\Z]^3$ using $\epsilon$-side length hypercubes for $\epsilon \geq 7/15$, and all $\epsilon \in \left[\frac{1}{r+1/(r^2+r+1)}, \frac{1}{r-1/(r^2-1)}\right)$.

For larger $n$, it is possible to prove $\mu_{n, \epsilon} = O(n \epsilon^{-n})$ using the probabilistic method \cite{bollobasCoveringTranslatesSet2011} (see also discussion in \cite{bogdanovCoveringThreeToriCubes2021}).
Thus, we have $\kappa_{n, d, \epsilon^d} \leq O(n\epsilon^{-n})$, and hence $\rho_{n, d, c} \leq O((n/c)^{d/n})$.

\subsection{\texorpdfstring{$n = 3$, $d = 2$}{n=3,d=2}}

Kopparty and Nagargoje studied the case of $n = 3$, and $d=2$ in \cite{KN}. 
\Cref{table:n3d2} contains a summary of their results.

\begin{table}[h]
    \begin{tabular}{l|l|l|l}
    \textbf{$c$} & \textbf{Best known partition}  & \textbf{$2$-d projection volumes} & \textbf{Lower bound} \\ \hline
    2            & Majority                       & $3/4$                             & $3/4$                \\
    3            & Golden Ratio                   & $1/\varphi \approx 0.618$            & $0.526$              \\
    \end{tabular}
    \caption{Partitions and lower bounds for $n=3$, and $d=2$ from \cite{KN}.}
    \label{table:n3d2}
\end{table}

Here is a description of the two partitions.

\textbf{Majority.} $f_{\text{Majority}}: \OI^3 \to \OI$ is the majority function on the discrete cube. Recall that functions on the discrete cube can be extended to partitions of $\OIS^n$ by rounding each coordinate.

\textbf{Golden Ratio Partition.}
Let $\psi = 1/\varphi$ where $\varphi$ is the golden ratio. Define the following partition $f_{GR}: \OIS^3 \to [3]$ as 
$$
f_{GR}(x, y, z) = \begin{cases}
    1, & |x| > \psi, |y| > \psi \\
    2, & |x| \leq \psi, |y| \leq \psi, z \leq 1/2 \\
    3, & \text{else.}\\
\end{cases}
$$

%
%





\section{Partition Sauer-Shelah}

We first show that there are partitions with non-trivial $d$-dimensional projections (non-trivial meaning projection volumes of $< 1$) whenever $n/c < d$. In contrast, it turns out that when $d$ is any smaller than $\frac{n}{c}$  we automatically have a $d$-dimensional projection which equals all of $[0,1]^d$ -- this is the (simple) partition analogue of the Sauer-Shelah lemma for this setting, which turns out to be tight.

We state and prove both these statements below.

\begin{lemma}[Partition Sauer-Shelah]\label{lem:CPSS}
    Let $n, d, c \in \Z^+$ with $d \leq \floor{n/c}$.
    Then, $\rho_{n,d,c} = 1$. 
    I.e., if $\mathcal{C}$ is $c$-cover of $\OIS^n$, there is some part $K \in \mathcal{C}$, and some subset $A \in \binom{[n]}{d}$ such that $|K_A| = 1$.
\end{lemma}

\begin{proof}
    Let $n, d, c \in \Z^+$ with $d \leq \floor{n/c}$.
    By contradiction, suppose the lemma was false, that is, there exists a $c$-cover $\mathcal{C} = \left\{ K^{(1)}, K^{(2)},..., K^{(c)} \right\}$ of $\OIS^n$ such that for all $K \in \mathcal{C}$, and $A \in \binom{[n]}{d}$, $K_A \neq \OIS^d$.

    Let $S_1,...,S_c \subset [n]$ such that $S_1,..,S_c$ are disjoint and each have size $d$.
    Then, for each $i \in c$, let $\alpha_i \in \OI^{S_i}$ such that $\alpha_i \notin K^{(i)}_{S_i}$.
    Then, let $\alpha \in \OI^n$ be some element that agrees with each $\alpha_i$. 

    Then $\alpha \notin K^{(i)}$ for any $i$, which contradicts the assumption that $\mathcal{C}$ covers $\OIS^n$.
\end{proof}

\begin{lemma}[Converse to the Partition Sauer-Shelah]\label{lem:PSS}
    For any $n, c, d \in \Z^+$, where $n / c <  d \leq n$, then there exists a $c$-cover $\mathcal{C}$ of $\OIS^n$ such that 
    $$
    \mpv{\mathcal{C}}{n, d} < 1 - \left( \frac{1}{c} \right)^{\ceil{n/c}}.
    $$
\end{lemma}

\begin{proof}
    Let $n, c, d \in \Z^+$ be such that  $n / c < d \leq n$. 
    Define 
    $$
    F^{(i)} = \left\{ \mathbf{x} \in \OIS^n: \left|\left\{j \in [n] : x_j \in \left[\frac{i-1}{c}, \frac{i}{c}\right]\right\}\right| \leq n/c \right\}.
    $$
    I.e. $F^{(i)}$ is the set of $\mathbf{x}$ such that $\leq n/c$ coordinates are in the interval $[(i-1)/c, i/c]$.
    Then, $\mathcal{C} = \left\{ F^{(1)},...,F^{(c)} \right\}$ cover $\OIS^n$.
    Indeed, for any $\mathbf{x} \in \OIS^n$, each of the $x_j$ are in at least one of the intervals $[(i-1)/c, i/c]$. 
    Hence, the average interval contains $n/c$ of the $x_j$; in particular, some interval has at most $n/c$ of them.

    We now bound $\mpv{\mathcal{C}}{n, d}$.
    Fix and $S \subset [n]$ with $|S| = d$.
    Then we have
    $$
    F^{(i)}_S \cap \left[\frac{i-1}{c}, \frac{i}{c}\right]^S = \emptyset,
    $$
    which implies 
    $$
    |F^{(i)}_S| \leq 1 - \left( \frac{1}{c} \right)^{d}.
    $$
\end{proof}

\section{Partitions from low influence functions}
\label{sec:inf}

In this section, we relate our problem to the well-studied notion of influence, which allows us to leverage known constructions and lower bounds.

\subsection{Projections of partitions and influences}\label{sec:inf-lowerbounds}

This subsection lower bounds $\mpv{f}{n,d}$ as a function of $\MaxInf_{n-d}(f)$ (\Cref{lem:influence-and-projection,cor:influence-and-projection}).
Thus, if our goal is to find partitions $f:\OIS^n \to [c]$ with small $d$-dimensional projection volumes, we should restrict our attention to those functions $f$ where groups of $n-d$ coordinates have small influence. 

Furthermore, we use this connection in conjunction with the BKKKL theorem to prove a lowerbound on $\rho_{n, n-1, 2}$ (\Cref{lem:lower-bound-c-2}).

\begin{lemma} \label{lem:influence-and-projection}
    For any function $f: \OIS^n \to [c]$, and $S \subset [n]$ such that $S \neq [n]$, if $\rho = \max_{\alpha \in [c]}(|f^{-1}(\alpha)_{S}|)$, and $\gamma = \Inf_f(\overline{S})$, then, $c\rho \geq 1 + \gamma$.
\end{lemma}

\begin{proof}
    Let $M(x, \alpha) = \1_{\{\alpha \text{ is above } x\}}$.
    Then we have 
    \begin{align*}
        \sum_{\alpha = 1}^c \int_{\OIS^S} M(x, \alpha) dx &= \int_{\OIS^S} \sum_{\alpha = 1}^c  M(x, \alpha) dx\implies \\
        \sum_{\alpha = 1}^c |f^{-1}(\alpha)_S| &= \int_{\OIS^S} |A(x, S)| dx\implies \\
    \end{align*}

    The LHS is at most $c\rho$, and the RHS is equal to $\E_{x \from \OIS^S}|A(x, S)|$, which is at least $1+\gamma$ (since $|A(x, S)|$ is 1 with probability $1 - \gamma$ and at least $2$ with probability $\gamma$). 
    The result follows.
\end{proof}

Taking the maximum over $S \subset \binom{[n]}{d}$, we get the following.
\begin{corollary}\label{cor:influence-and-projection}
    For any $n, d, c \in \Z^+$, and $f: \OIS^n \to [c]$, $\mpv{f}{n, d, c} \geq \frac{1 + \MaxInf_{n-d}(f)}{c}$.
\end{corollary}

From this, we also see that lower bounds on influence translate to lower bounds on the maximum projection size.


In the case of $c = 2$, and $d = n - 1$, we can apply this to the BKKKL theorem to get the following.

\begin{lemma}\label{lem:lower-bound-c-2}
    For any $n$, $\rho_{n, n-1, 2} \geq 1/2(1 + \Omega(\log n / n)))$.
\end{lemma}

\begin{proof}
    Let $f$ be any cover of $\OIS^n$ into $2$ parts. 
    Suppose $\Pr(f(x) = 1) = 1/2 + \epsilon$ for some $\epsilon \in [0,1/2]$.
    Then, by \Cref{thm:bkkkl}, we have that there exists some coordinate with influence $\Omega((1/4 - \epsilon^2)\log(n) / n)$.
    Applying this to \Cref{cor:influence-and-projection}, we have that 
    $$
    \mpv{f}{n, n-1, 2} \geq \frac{1}{2} (1 + \Omega((1/4 - \epsilon^2)\log(n) / n)).
    $$

    Additionally, using the volume lower bound  (\Cref{lem:volume-lower-bound}) to $f^{-1}(1)$, we get
    $$
    \mpv{f}{n, n-1, 2} \geq (1 / 2 + \epsilon)^{(n-1)/n} \geq \frac{1}{2} + \epsilon  - \frac{(\frac{1}{2} + \epsilon) \log(\frac{1}{2} + \epsilon)}{n} .
    $$

    If, $\epsilon < 0.01$, then the first bound implies $\mpv{f}{n, n-1, 2} \geq \frac{1}{2}(1 + \Omega(\log(n) / n)) $, and if $\epsilon > 0.01$, then the second bound implies $\mpv{f}{n, n-1, 2} \geq \frac{1}{2}(1 + 2\epsilon - o(1)) = \frac{1}{2}(1 + \Omega(\log(n) / n))$.
\end{proof}

Compare this to the general lower bound, which only gives $\rho_{n, n-1, 2} \geq 1/2 + \Omega(\frac{1}{n})$.

\subsection{Main Result: Partitions with small projections from low influence functions}\label{sec:inf-upperbounds}

In this subsection we construct partitions of $\OIS^n$ with small $n-1$ dimensional projections from low influence functions. 
We first show that the Tribes function provides an asymptotically optimal partition for $c=2$. 
Next we show how to obtain small shadow partitions for $c > 2$ parts, culminating in the following theorem.

\begin{theorem}\label{thm:main}
    For any $g(n) = o(n)$, there exists infinitely many $n$ such that there exists a partition, $f: \OI^n \to [c]$, where $c = 2^{g(n)}$,  with $\mpv{f}{n, n-1, c} \leq \frac{1}{c}(1 + o(1))$.
\end{theorem}

Note that for $c = 2^{o(n)}$, this is asymptotically tight for the general lowerbound (\Cref{lem:general-lower-bound}), which gives $\mpv{f}{n, n-1, c} \geq \frac{1}{c}(1 + o(1))$.
This improves upon the work \cite{KN} who give partitions with projection volume $\frac{1}{c}(1 + o(1))$ up to $c = O(\sqrt{n})$.

We start by showing a converse of \Cref{cor:influence-and-projection} (i.e., functions with small influence have small maximum projection sizes).
We are able to show this for functions $f$ from the discrete cube $\OI^n$.
Recall that one can obtain $c$-partitions of $\OIS^n$ from $f:\OI^n \to [c]$ by rounding each coordinate.

\begin{lemma}[Paritions from low influence boolean functions]\label{lem:partitions-from-boolean-valued-functions}
    Let $f: \OI^n \to [c]$ be an $\epsilon$-balanced function with $\MaxInf(f) = \gamma$. Then, $\mpv{f}{n, n-1, c} \leq \frac{1}{c}\left( 1 + \frac{c\gamma}{2} + c \epsilon\right)$.
\end{lemma}

\begin{proof}
    Let $f:\OI^n \to \OI$ be an $\epsilon$-balanced boolean function, and let $S = [n] \setminus i$ for some $i \in [n]$.

    For any $\alpha \in [c]$, let $A_{\alpha} = \{x \in \OI^S: \alpha \in A(x, S) \text{ and } |A(x, S)| = 2\}$, and $B_{\alpha} = \{x \in \OI^S: A(x, S) = \left\{ \alpha \right\}\}$.
    In words, $A_\alpha$ contains all of the $x \in \OI^S$ such that $\alpha$ is above $x$, and some other color is also above $x$, and $B_\alpha$ contains all of the $x \in \OI^S$ such that only $\alpha$ is above $x$.
    Notice that $A_\alpha$, and $B_\alpha$ are disjoint and that their union is exactly $f^{-1}(\alpha)_S$. 
    Hence 
    $$
        |f^{-1}(\alpha)_S| = |A_\alpha| + |B_\alpha|.
    $$
    On the other hand, we have
    $$
        |f^{-1}(\alpha)| = \frac{1}{2}|A_\alpha| + |B_\alpha|.
    $$
    One way to see this is to sample $y \in \OI^n$ by sampling $x \in \OI^S$ and then picking a random binary value for the $i$th coordinate. 
    If $x \in B_\alpha$, then $f(y) = \alpha$ with probability 1, and if $x \in A_\alpha$, $f(y) = \alpha$ with probability $1/2$.
    
    Combining the two equations, we get $|f^{-1}(\alpha)_S| = |f^{-1}(\alpha)| + \frac{1}{2}|A_\alpha|.$

    We now bound $|A_\alpha|$. 
    We have $\sum_{\alpha \in [c]} |A_\alpha| = 2\gamma$ since each $x$ for which the $i$th coordinate has influence is in exactly two of the $A_\alpha$. 
    Furthermore, any particular $A_\alpha$ has size at most $\gamma$ since every $x$ that contributes to $A_\alpha$ also contributes to some other $A_i$. 
    Thus, $|A_\alpha| \leq \gamma$, and we have

    $$
    |f^{-1}(\alpha)_S| \leq |f^{-1}(\alpha)| + \frac{\gamma}{2} \leq \frac{1}{c} + \epsilon + \frac{\gamma}{2},
    $$
    where we got the last inequality by the assumption that $f$ is $\epsilon$-balanced.
\end{proof}

Thus, balanced functions $f:\OI^n \to [c]$ with low maximum influence give upperbounds on $\rho_{n, n-1, c}$.
\begin{corollary}\label{cor:influence-to-rho}
    If there is an $\epsilon$-balanced boolean function $f: \OI^n \to [c]$ with maximum influence at most $\gamma$, then 
    $\rho_{n, n-1, c} \leq \frac{1}{c}\left( 1 + \frac{c\gamma}{2} + c \epsilon \right)$.
\end{corollary}

In the case $c=2$, applying this to directly to the Tribes function yields the following.

\begin{corollary} \label{cor:partition-from-tribes}
    $\rho_{n, n-1, 2} \leq 1/2 + O\left(\frac{\log(n)}{n}\right)$.
\end{corollary}
Note that this is tight when compared to the lower bound in the previous section (\Cref{lem:lower-bound-c-2}).
Next, we lift this partition for $c > 2$ by taking the `product partition'

\begin{definition}[Product of Partitions]
    Let $f_1: \OI^{n_1} \to [c_1]$, $f_2: \OI^{n_2} \to [c_2]$ be covers of $\OI^{n_1}$ and $\OI^{n_2}$ respectively. Define their product $f_1 \times f_2 : \OI^{n_1} \times \OI^{n_2} \to [c_1] \times [c_2]$ as $f_1 \times f_2(x, y) = (f_1(x), f_2(y))$, and view this as a cover of $\OI^{n_1 + n_2}$ into $c_1c_2$ parts.
\end{definition}

The next lemma shows that the product partition of good partitions is good.
\begin{lemma}\label{lem:product-of-partitions}
    Let $n_1, n_2, c_1, c_2 \in \N$, and $n = n_1 + n_2$, and $c = c_1\cdot c_2$. 
    Let $\epsilon_1, \epsilon_2 \in (0,1)$, and $f_1: \OI^{n_1} \to [c_1]$, $f_2: \OI^{n_2} \to [c_2]$ be $\epsilon_1$ and $\epsilon_2$-balanced covers of $\OI^{n_1}$ and $\OI^{n_2}$ respectively.
    Then $f = f_1 \times f_2$ is a cover of $\OI^{n}$ into $c$ colors such that 
    \begin{enumerate}[1.)]
        \item $\Pr[f = c] \leq (1/c_1 + \epsilon_1)(1/c_2 + \epsilon_2)$
        \item $\mpv{f}{n, n-1} = \max\left\{\mpv{f_2}{n_2, n_2-1} \cdot (1/c_1 + \epsilon_1), \mpv{f_1}{n_1, n-1}\cdot (1/c_2 + \epsilon_2)\right\}$
    \end{enumerate}
\end{lemma}

\begin{proof}
    The first item is obvious.

    For the second item, let $i \in [n]$, and let $(a, b)$ be an arbitrary element of $[c_1] \times [c_2]$. 
    If $i \leq n_1$, then 
    \begin{align*}
        |f^{-1}((a, b))_{[n]\setminus i}| &=\Pr_{x \in \OI^{[n] \setminus \left\{ i \right\}}}((a, b) \text{ is above }x) \\
                             &=\Pr_{x_1 \in \OI^{[n_1] \setminus \{i\}}}(a \text{ is above } x_1) \cdot \Pr_{x_2 \in \OI^{[n_2]}}(b \text{ is above }x_2)\\
                             &\leq| f_1^{-1}(a)_{[n_1] \setminus \left\{ i \right\}}|\cdot(1/c_2 + \epsilon_2)\\
                             &\leq\mpv{f_1}{n_1, n_1 - 1}\cdot(1/c_2 + \epsilon_2)\\
    \end{align*}
    If $i > n_1$, the argument is symmetric.
\end{proof}

Thus, we see that taking the product of two balanced partitions with small projection sizes yields another partition with small projection sizes and more colors.
When taking the product of a cover with itself many times, we get the following corollary.

\begin{corollary}\label{cor:product-many-times}
    Let $f: \OI^{n} \to [c]$ be a $\epsilon$-balanced cover of $\OI^n$ into $c$ parts. Applying the previous claim inductively, we get that $f^k$ is a cover of $\OI^{kn}$ into $c^k$ parts, with
    $\mpv{f^k}{kn,n-1} \leq \mpv{f}{n, n-1} \cdot (1/c + \epsilon)^{k-1}$
\end{corollary}


Finally, applying this to the majority function, we get the following.

\begin{corollary}[Product of Majority]\label{cor:product-of-majority}
    For any $n, k$, where $n$ is divisible by $k$, $(\Maj_{n/k})^k$ is a function from $\OI^n \to [c]$, where $c=2^k$ such that 
    $$\mpv{(\Maj_{n/k})^k}{n, n-1} \leq \frac{1}{c}(1 + O(\sqrt{k/n})).$$
\end{corollary}

To get the \Cref{thm:main}, we simply note that for $k=o(n)$, $\sqrt{k/n} = o(1)$.

We can also apply \Cref{cor:product-many-times} to the tribes function to get better quantitatives.
\begin{corollary}[Product of Tribes]\label{cor:product-of-tribes}
    There exists infinitely many $n, k$ where $n$ is divisible by $k$ such that $(\Tribes_{n/k})^k$ is a function from $[0,1]^n \to [c]$, where $c=2^k$ such that 
    $$\mpv{f}{n, n-1} \leq \frac{1}{c}(1 + O(\frac{k}{n}\log\frac{n}{k})).$$
\end{corollary}

%

\subsection{\texorpdfstring{$c = 2^{b n}$}{c = 2bn}}

This section explores the setting of $c = 2^{bn}$, where $b$ is a constant and makes a clean conjecture about $\rho_{n, n-1, 2^{bn}}$.

Applying the Product of Tribes construction (\Cref{cor:product-of-tribes}) to this setting (taking the product of $bn$ tribes of size $1/b$ each), we get a maximum projection volume of 
\[
    \frac{1}{c}(1 + O(b \log( 1 /b))),
\]
where the big-O notation assumes $b$ is going to zero.

On the other hand, the general lower bound (\Cref{lem:general-lower-bound}) is $\frac{1}{c}(1 + \Omega(b))$. 
We conjecture that it is impossible to do substantially better than the Product of Tribes’ construction in general.
Formally, we conjecture the following.

\begin{conjecture}\label{conj:lower-bound-inf}
    Let $b \in (0,1)$, and let $c = 2^{bn}$. Then, there exists a universal constant $\delta$ such that 
    $$\rho_{n, n-1, c} \geq \frac{1}{c}(1 + \delta b\log(1/b)).$$
    In other words, for any cover of $\OIS^n$ of size $2^{bn}$, there is some part of the cover, $K$, and some set of coordinates $A \subset [n]$ of size $n-1$, such that $|K_A| \geq \frac{1}{c}(1 + \delta b\log(1/b))$.
\end{conjecture}

Note that this conjecture being true would imply the KKL theorem by the following argument. If there exists a balanced boolean-valued function $g:\OI^n \to \OI$ with maximum influence $\gamma_n = o(\log(n) / n)$ (i.e. KKL was false), then the product of $bn$ instances $g$, each of size $1/b$, would yield a function $f$ from $\OI^{n} \to [2^{bn}]$ with $\mpv{f}{n,n-1} = 1/{c}(1 + o(b\log(1/b))$, contradicting the conjecture.

\printbibliography

\appendix
\section{Proof of the \texorpdfstring{$d=1$}{d=1} lower bound}\label{appendix:d1-lower}

This section proves the lower bound for $d = 1$ using the argument from \cite{mcelieceCoveringToriSquares1973}.
We restate the theorem here for convenience.
\begin{theorem*}
    Let $n, c$ be positive integers with $n\geq 1$, and let $\epsilon > 0$.
    If $\mathcal{C}$ is a $c$-cover of $\OIS^n$ with $\mpv{\mathcal{C}}{n,1} \leq \epsilon$, then $c \geq \ceil{\epsilon^{-1}}^{(n)}$.
    In other words, 
    $\kappa_{n, 1, \epsilon} \geq \ceil{\epsilon^{-1}}^{(n)}$.
\end{theorem*}

\begin{proof}
    By induction on $n$. 

    \textbf{Base case $n=1$.} Let $\mathcal{C}$ be a $c$-cover of the interval $\OIS$ with $\mpv{\mathcal{C}}{n, 1} \leq \epsilon$. 
    Then, the total length of any part of $\mathcal{C}$ has length at most $\epsilon$. 
    Thus, $c \geq \epsilon^{-1}$. Since the number of parts is integral, we need at least $c \geq \ceil{\epsilon^{-1}}$, as required.

    \textbf{Inductive step.} For the inductive step, let $n \geq 2$, and suppose the claim is true for $n - 1$. 
    We'll show the claim is true for $n$. 
    Let $\mathcal{C}$ be any $c$-cover of $\OI^n$ with $\mpv{\mathcal{C}}{n, 1} \leq \epsilon$. 

    For each part $K \in \mathcal{C}$, let $B(K) \subset \OIS^n$  be a set satisfying the following conditions.
    \begin{enumerate}[1.)]
        \item $K \subset B(K)$, and 
        \item $B(K)$ is a product set of intervals of length exactly $\epsilon$.
    \end{enumerate}

    One can find $B(K)$ by projecting $K$ onto each coordinate axis, extending the projects to have length exactly $\epsilon$, and then taking their product.
    Note that this is always possible since all one-dimensional projections of $K$ have length at most $\epsilon$.

    Let $\mathcal{D} = \left\{ B(K): K \in \mathcal{C} \right\}$.

    We will establish the bound by computing the total volume of parts in $\mathcal{D}$, $V$, in two ways.

    Note 
    $$V = \sum_{K \in \mathcal{D}} |K|,$$ 
    and since each $K \in \mathcal{D}$ has volume exactly $\epsilon^n$, we have $V = c \cdot \epsilon^n$. 

    We now calculate $V$ in another way to leverage the inductive hypothesis.
    Define $K_{a} = \{(x_2,...,x_n): \mathbf{x} \in K \text{ and } x_1 = a\}$.
    Note that for any $K \in \mathcal{D}$, $a \in \OIS$, $K_a$ is either a product of $n-1$ intervals of length exactly $\epsilon$, or the empty set.
    Let $\mathcal{D}_a = \left\{ K_a: K \in \mathcal{D}, K_a \neq \emptyset \right\}$.
    Since $\mathcal{D}$ is a cover of $\OIS^n$, $\mathcal{D}_a$ is a cover of $\OIS^{n-1}$.
    Furthermore, $\mpv{\mathcal{D}_a}{n,1} \leq \epsilon$, so the inductive hypothesis applies, and we have $|\mathcal{D}_a| \geq \ceil{\epsilon^{-1}}^{(n-1)}$.

    Then,
    \begin{align*}
        V &= \sum_{K \in \mathcal{D}} \int_{[0,1]} |K_a| da\\
          &= \int_{[0,1]}\sum_{K \in \mathcal{D}}  |K_a| da\\
          &= \int_{[0,1]}\sum_{K \in \mathcal{D}_a}  \epsilon^{n-1} da\\
          &\geq \int_{[0,1]} \ceil{\epsilon^{-1}}^{(n-1)}\epsilon^{n-1}  da \\
          &= \epsilon^{n-1}\ceil{\epsilon^{-1}}^{(n-1)}.\\
    \end{align*}

    Thus, we have 
    \begin{align*}
        c \epsilon^n &\geq \epsilon^{n-1}\ceil{\epsilon^{-1}}^{(n-1)} \implies \\
        c &\geq \epsilon^{-1}\ceil{\epsilon^{-1}}^{(n-1)}\implies \\
        c &\geq \ceil{\epsilon^{-1}}^{(n)},
    \end{align*}
    where the last inequality is because $c$ is an integer.
\end{proof}

\end{document}